\documentclass[twocolumn,superscriptaddress,showpacs,pra]{revtex4}
\usepackage{epsfig}
\usepackage{epstopdf}
\usepackage{delarray}
\usepackage{amsmath, amssymb}
\usepackage{bm}
\usepackage{graphicx}% Include figure files
\usepackage{placeins}
\usepackage{color}
\begin{document}
\title{Self-Localized Solutions of the Kundu-Eckhaus Equation in Nonlinear Waveguides}

\author{Cihan Bay\i nd\i r}
\email{cihanbayindir@gmail.com}
\affiliation{Associate Professor, Engineering Faculty, \.{I}stanbul Technical University, 34467 Maslak, \.{I}stanbul, Turkey. \\
						 Adjunct Professor, Engineering Faculty, Bo\u{g}azi\c{c}i University, 34342 Bebek, \.{I}stanbul, Turkey. \\
						 International Collaboration Board Member, CERN, CH-1211 Geneva 23, Switzerland.}

%\date{\today}
\begin{abstract}
In this paper we numerically analyze the 1D self-localized solutions of the Kundu-Eckhaus equation (KEE) in nonlinear waveguides using the spectral renormalization method (SRM) and compare our findings with those solutions of the nonlinear Schr\"{o}dinger equation (NLSE). For cubic-quintic-Raman nonlinearity, as a benchmark problem we numerically construct single, dual and $N$-soliton solutions for the zero optical potential, i.e. $V=0$, which are analytically derived before. We show that self-localized soliton solutions of the KEE with cubic-quintic-Raman nonlinearity do exist, at least for a range of parameters, for the photorefractive lattices with optical potentials in the form of $V=I_o \cos^2(x)$. Additionally, we also show that self-localized soliton solutions of the KEE with saturable cubic-quintic-Raman nonlinearity do also exist for some range of parameters. However, for all of the cases considered, these self-localized solitons are found to be unstable. We compare our findings for the KEE with their NLSE analogs and discuss our results.

\end{abstract}
\maketitle

%%%%%%%%%%%%%%%%%%%%%%%%%%%%%%% main %%%%%%%%%%%%%%%%%%%%%%%%%%%%%
\section{\label{sec:level1} Introduction}
The study of nonlinear wave equations appear everywhere in applied mathematics and theoretical physics including engineering and bio-sciences. These equations provide good examples of dynamical systems which possess diverse phenomena. The list of this diverse phenomena include but are not limited to solitary waves, rogue waves, the formation of singularities, dispersive turbulence and the propagation of chaos, just to name a few.  Nonlinear waves occur in physical and natural systems and the studies of nonlinear optics and fiber optics, water and atmospheric waves, and turbulence in hydrodynamics and plasmas represent their important applications. 

Some well-known nonlinear partial differential equations are the Korteweg-de Vries equation, nonlinear Klein-Gordon equation, nonlinear Schrödinger equation (NLSE) etc. In this paper we study the self-localized solutions of the Kundu-Eckhaus equation (KEE), which is an NLSE like equation. Widely accepted form of the KEE has two additional terms compared to the cubic NLSE. These terms are the quintic nonlinear term which accounts for the higher order nonlinearity and the Raman-effect term which accounts for the self-frequency shift of the waves.

KEE equation can adequately model the propagation of ultrashort pulses in nonlinear and quantum optics, which can possibly be used to describe the optical properties of the femtosecond lasers and can be used in femtochemistry studies. In mechanics, KEE is capable of examining the stability of Stokes waves in weakly nonlinear dispersive media. In plasma physics it can be used to model ion-acoustic waves. Some extensions of the NLSE, similar to the form of the KEE, where quintic nonlinearity is not included but third order dispersion and gain and loss terms are included are also used as models in the soliton-similariton laser studies \cite{Geng, Wang_KeeNsoliton, BayPRE1, BayPRE2, Wang, Zhao2013, dqiu}. Some analytical solutions of the KEE, including self-localized solutions of $sech$ type, are analytically obtained by utilizing different techniques such as Darboux and Backlund transformations, the first integral method and exp-function method \cite{Geng, Wang, Zhao2013, dqiu, KilicKEEsech, Wanga}. While single and dual self-localized solitons are derived two decades ago \cite{Geng}, the N-soliton solutions of the KEE are derived recently \cite{Wang_KeeNsoliton, Wanga}.

With these motivations, in this paper we study the self-localized solutions of the KEE numerically. For this purpose we implement spectral renormalization scheme to investigate the self-localized solutions of the KEE. For the cubic nonlinearity, we numerically derive analytically derived single, dual and N-solution solutions as a benchmark problem and then extend the analysis to periodic potentials. Then we focus on saturable nonlinearity and show that self-localized solutions of the KEE can exist for some range of parameters involved. We compare our findings with their NLSE counterparts and discuss various aspects of our results.

\section{\label{sec:level2}Methodology}

\subsection{Spectral Renormalization Method for the Kundu-Eckhaus Equation} 

Self-localized solutions of many nonlinear systems can be found by different computational techniques. These include but are not limited to shooting, self-consistency and relaxation \cite{Ablowitz, Bay_CSRM}. One of the most popular methods is the Petviashvili's method. In Petviashvili's method, the governing nonlinear equation is transformed into Fourier space as in the case of general Fourier spectral schemes \cite{Canuto, Karjadi2010,Karjadi2012, trefethen,  BayPRE1, BayPRE2, BayTWMS2016, Demiray2015, BayPLA, Baysci, BayTWMS2015, Bay_arxNoisyTun, Bay_arxNoisyTunKEE, Bay_arxEarlyDetectCS}, and  a convergence factor is determined according to the degree of the nonlinear term \cite{Petviashvili, Ablowitz}. This method was first introduced by Petviashvili and applied to the Kadomtsev-Petviashvili equation \cite{Petviashvili}. Later, it has been applied to many other systems for modeling many different phenomena such as dark and gray solitons and lattice vortices, just to name a few \cite{Ablowitz, Yang}. Petviashvili's method works well for nonlinearities with fixed homogeneity only therefore this method is extended to spectral renormalization method (SRM), which is can be used to find the localized solutions in waveguides with other types of nonlinearities \cite{Ablowitz, Fibich}. Later another extension which is known as compressive spectral renormalization method (CSRM) is proposed \cite{Bay_CSRM}, in order to obtain stable self-localized solutions in nonlinear waveguides with missing spectral data.

The SRM essentially transforms the governing equation into wavenumber space by means of Fourier transform and couples it to a nonlinear integral equation. This nonlinear integral equation is basically an energy conservation principle used in the iterations in the wavenumber space \cite{Ablowitz}. Due to this coupling, the initial conditions converge to the self-localized solutions of the nonlinear system modeled \cite{Ablowitz}. SRM is efficient, is easy to implement and it can be applied to many different dynamic nonlinear models with different higher-order nonlinearities \cite{Ablowitz}. In this section we apply the SRM to the KEE to obtain its self-localized solutions in waveguides. We start with the KEE in the form of
 
\begin{equation}
i\zeta_z +  \zeta_{xx}+2 \left| \zeta \right|^2 \zeta+ \beta^2 \left| \zeta \right|^4 \zeta-2 \beta i  \left( \left| \zeta \right|^2 \right)_x \zeta -V(x)\zeta=0
\label{eqin01}
\end{equation}
where $z$ is the propagation direction of optical pulse, $x$ is the transverse coordinate, $i$ denotes the imaginary unity, $\beta$ is a real constant which controls the skewness of the solutions and $\beta^2$ is the coefficient of the quintic nonlinear term which model the effects of higher order nonlinearity \cite{BayPRE1, BayPRE2, Wang, Zhao2013, dqiu}. The last term models the Raman effect accounting for the self-frequency shift of the pulses and $\zeta$ is complex amplitude of the optical field. Eq.~(\ref{eqin01}) can be rewritten as
\begin{equation}
i\zeta_z +  \zeta_{xx} -V(x)\zeta+ N(\left| \zeta \right|^2) \zeta =0
\label{eq01}
\end{equation}
where $N(\left| \zeta \right|^2)=2 \left| \zeta \right|^2+\beta^2 \left| \zeta \right|^4-2 \beta i  \left( \left| \zeta \right|^2 \right)_x $ . Using the ansatz, $\zeta(x,z)=\eta(x,\mu) \textnormal{exp}(i\mu z)$, where $\mu$ shows the soliton eigenvalue, the KEE becomes
\begin{equation}
-\mu \eta +  \eta_{xx} -V(x)\eta+ N(\left| \eta \right|^2) \eta =0
\label{eq02}
\end{equation}
Furthermore taking the 1D Fourier transform of $\eta$ one can obtain
\begin{equation}
\widehat{\eta} (k)=F[\eta(x)] = \int_{-\infty}^{+\infty} \eta(x) \exp[i(kx)]dx
\label{eq03}
\end{equation}
For a zero optical potential, $V=0$, the Fourier transform of Eq.~(\ref{eq02}) in 1D yields to
\begin{equation}
\widehat{\eta} (k)=\frac{F \left[ N( \left| \eta \right|^2\eta) \right]}{\mu+\left| k \right|^2}
\label{eq04}
\end{equation}
The formula given in Eq.~(\ref{eq04}) may be applied iteratively to find the self-localized solutions of the model equation. This procedure was proposed by Petviashvili in \cite{Petviashvili} for the first time. However the iterations of Eq.~(\ref{eq04}) may grow unboundedly or it may tend to zero as discussed in \cite{Ablowitz}. This problem can be addressed by introducing a new variable in the form $\eta(x)=\alpha \xi(x)$. The 1D Fourier transform of this new variable reads $\widehat{\eta}(k)=\alpha \widehat{\xi}(k)$. Using these substitutions, Eq.~(\ref{eq04}) becomes
\begin{equation}
\widehat{\xi} (k)=\frac{F\left[ N( \left| \alpha \right|^2 \left| \xi \right|^2 ) \xi \right]}{\mu+\left| k \right|^2}=R_{\alpha}[\widehat{\xi} (k)]
\label{eq05}
\end{equation}
and corresponding iteration scheme can be written as
\begin{equation}
\widehat{\xi}_{j+1} (k)=\frac{F\left[N( \left| \alpha_j \right|^2 \left| \xi_j \right|^2 ) \xi_j \right]}{\mu+\left| k \right|^2}
\label{eq06}
\end{equation}
An algebraic condition on the parameter $\alpha $ can be obtained using the energy conservation principle for the normalization part of the SRM. By multiplying both sides of Eq.~(\ref{eq05}) with the $\widehat{\xi}^*(k)$, which is the complex conjugate of $\widehat{\xi}(k)$, and integrating to evaluate the total energy, one can obtain the algebraic condition as 
\begin{equation}
\int_{-\infty}^{+\infty} \left|\widehat{\xi} (k)\right|^2 dk= \int_{-\infty}^{+\infty} \widehat{\xi}^* (k) R_{\alpha}[\widehat{\xi} (k)]dk  
\label{eq07}
\end{equation}
which is the normalization constraint. This constraint guarantees that the scheme converges to a self-localized soliton. The procedure of obtaining self-localized solutions of a nonlinear system, which is applied to KEE in this paper, summarized above is known as the SRM \cite{Ablowitz}. Using a single or multi-Gaussians as initial conditions, Eq.~(\ref{eq04}) and the normalization constraint given in Eq.~(\ref{eq07}) are applied iteratively to find the profile for each iteration count. Iterations are continued until the parameter ${\alpha}$ convergences.  

Returning to a more general setting, the nonzero potentials ($V\neq0$) are widely used as models for various optical media i.e.  nondefected or defected photonic crystals. To avoid singularity of the scheme, one can add and substract a $p \eta$ term with $p>0$ from Eq.~(\ref{eq02}) \cite{Ablowitz}. Then the 1D Fourier transform of Eq.~(\ref{eq02}) becomes
\begin{equation}
\widehat{\eta} (k)=\frac{(p+| \mu|)\widehat{\eta}}{p+\left| k \right|^2} -\frac{F[V \eta]-F \left[ N(\left| \eta \right|^2) \eta \right]}{p+\left| k \right|^2}
\label{eq08}
\end{equation}
which is the iteration scheme for the KEE with a nonzero optical potential. In this paper we are specifically interested in photorefractive solitons of the KEE which are of practical use. Therefore, considering the 1D versions of the photorefractive solitons of the NLSE like equation first reported in \cite{Segev}, we set the optical potential as $V=I_o \cos^2(x)$ and the nonlinear term as $N(\left| \eta \right|^2)=-1/(1+2 \left| \eta \right|^2)$ for the saturable cubic nonlinearity and $N(\left| \eta \right|^2)=-1/(1+2 \left| \eta \right|^2+\beta^2 \left| \eta \right|^4-2 \beta i  \left( \left| \eta \right|^2 \right)_x)$ for the saturable cubic-quintic-Raman nonlinearity. Various forms of the saturable nonlinearities account for different field-induced changes in the refractive index \cite{Gatz}. Such saturation behavior, which determines the field strength when saturation occurs, can be modeled by considering the various particular physical processes involved \cite{Gatz}. Compared to the cubic NLSE, the form of the saturable nonlinearity used in this paper can be utilized to model the field particularly under the effect of quintic nonlinearity and Raman scattering phenomena. As before, we can define a new parameter $\eta(x)=\alpha \xi(x)$ and find its Fourier transform as $\widehat{\eta}(k)=\alpha \widehat{\xi}(k)$. Using these substitutions iteration formula for saturable nonlinearity becomes
\begin{equation}
\begin{split}
&\widehat{\xi}_{j+1} (k) =\frac{(p+| \mu|)}{p+\left| k \right|^2}\widehat{\xi_j}-\frac{F[V \xi_j]}{p+\left| k \right|^2} \\
& +\frac{1}{p+\left| k \right|^2}. \\ 
& F\left[ \frac{\xi_j}{1+2 \left| \alpha_j \right|^2 \left| \xi_j \right|^2+ \beta^2 \left| \alpha_j \right|^4 \left| \xi_j \right|^4- 2 \beta i . F^{-1}[ik F[\left| \alpha_j \right|^2 \left| \xi_j \right|^2]] }\right]  \\
& =R_{\alpha_j}[\widehat{\xi}_j (k)]
\label{eq09}
\end{split}
\end{equation}
The algebraic condition SRM for nonzero potentials can be attained by multiplying both sides of Eq.~(\ref{eq09}) with  $\widehat{\xi}^*(k)$ and integrating to evaluate the total energy, which leads to the normalization constraint as
\begin{equation}
\int_{-\infty}^{+\infty} \left|\widehat{\xi} (k)\right|^2 dk= \int_{-\infty}^{+\infty} \widehat{\xi}^* (k) R_{\alpha}[\widehat{\xi} (k)]dk  
\label{eq10}
\end{equation}
As in the case of zero potentials, an initial condition in the form of a single or multi-Gaussians converges to self-localized states of the model equation when Eq.~(\ref{eq09}) and Eq.~(\ref{eq10}) are applied iteratively. Iterations can be performed until the parameter ${\alpha}$ converges with a specified upper error bound. The reader is referred to \cite{Ablowitz} for a more comprehensive discussion and application of SRM to NLSE like equation and to second-harmonic generation analysis. We present our results for the KEE in the next section.

\section{\label{sec:level3}Results and Discussion}
 
\subsection{Self-Localized Soliton Solutions of the KEE for Zero Optical Potential}
%\begin{doublespace}
First we concentrate on the KEE with regular cubic-quintic-Raman nonlinearity with no optical potential, which can be obtained by setting $V=0$. The nonlinear term for this case is taken as $N(\left| \zeta \right|^2)=2 \left| \zeta \right|^2+\beta^2 \left| \zeta \right|^4-2 \beta i  \left( \left| \zeta \right|^2 \right)_x $ in Eq.~(\ref{eq01}). The parameters of the computations are selected as $p=10, \mu=1, I_0=0.1$. 

\begin{figure}[htb!]
\begin{center}
   \includegraphics[width=3.4in]{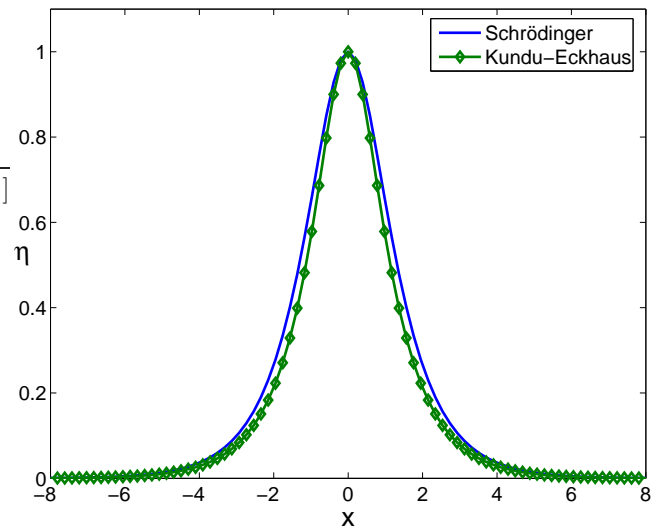}
  \end{center}
\caption{\small Comparison of the self-localized single soliton solutions of the KEE and NLSE.}
  \label{fig1}
\end{figure}

In Figure~\ref{fig1}, we compare the single humped self-localized soliton solution of the KEE with its NLSE counterpart. For this numerical solution we use $N=2048$ spectral components and define the convergence as the normalized change of $\alpha$ to be less than $10^{-15}$. The initial condition for this simulation is simply a Gaussian in the form of $\exp{(-(x-x_0)^2)}$ where $x_0$ is taken as $0$. SRM converges to the exact single sech type soliton solution analytically derived in \cite{Geng} within few iteration steps. The self-localized soliton solution of the KEE is more slender, that is its peak value is achieved for a narrower profile which can be measured using -3dB mainlobe width, compared to its NLSE counterpart. This is expected since the quintic nonlinear term in the KEE introduces higher order nonlinearity of the solutions.

\begin{figure}[htb!]
\begin{center}
   \includegraphics[width=3.4in]{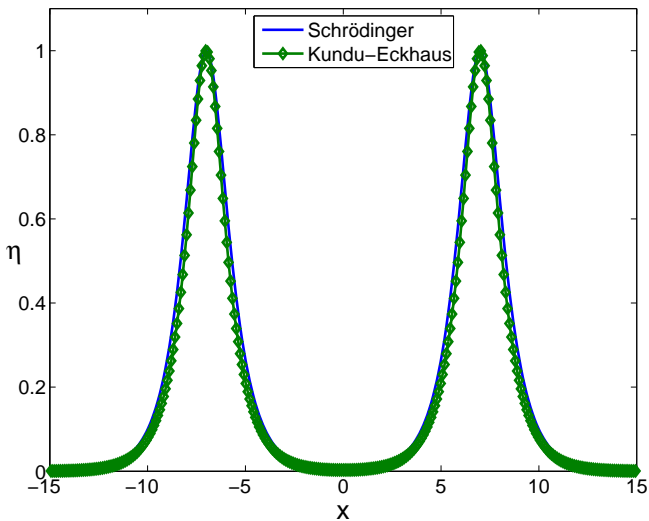}
  \end{center}
\caption{\small Comparison of the self-localized dual soliton solutions of the KEE and NLSE.}
  \label{fig2}
\end{figure}

In Figure~\ref{fig2}, the dual humped self-localized soliton solution of the KEE with its NLSE counterpart are compared. Again, for this numerical solution we use $N=2048$ spectral components. The convergence is defined as the normalized change of $\alpha$ to be less than $10^{-15}$ as before. The initial condition for this simulation is simply two Gaussians in the form of $\exp{(-(x-x_0)^2)}+\exp{(-(x-x_1)^2)}$ where $-x_0=x_1=10$. SRM converges to the dual sech type soliton solution within few iteration steps for which the analytical form of solutions are given in \cite{Geng, Wang_KeeNsoliton, Wanga}. The self-localized dual soliton solution of the KEE is again more slender, that is its peak value is achieved for a narrower profile which can be measured using -3 dB mainlobe width, compared to its NLSE counterpart for which analytical expressions are also known. Again, this is an expected result due to higher order nonlinearity of the KEE compared to the NLSE.

\begin{figure}[htb!]
\begin{center}
   \includegraphics[width=3.4in]{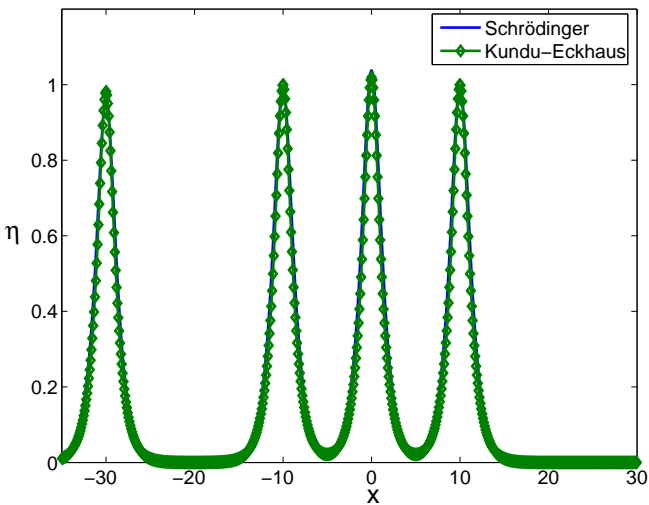}
  \end{center}
\caption{\small Comparison of the self-localized N-soliton solutions of the KEE and NLSE.}
  \label{fig3}
\end{figure}

In Figure~\ref{fig3}, the self-localized soliton solution of the KEE with 4 humps with its NLSE counterpart are compared. Same number of spectral components are used as before. The convergence is defined as the normalized change of $\alpha$ to be less than $10^{-7}$ for this simulation. The initial condition for this simulation is simply four Gaussians in the form of $\exp{(-(x-x_0)^2)}+\exp{(-(x-x_1)^2)}+\exp{(-(x-x_2)^2)}+\exp{(-(x-x_3)^2)}$ where $x_0=0, x_1=10, x_2=-10, x_3=-30$. SRM converges to the sech type soliton solution with 4 humps within few iteration steps. The self-localized soliton solution of the KEE for this case is more slender as before. If we add more Gaussians (i.e. N-Gaussians) in the initial condition, we observe that SRM converges to construct the self-localized N-soliton solutions for which the analytical solutions are given in \cite{Wang_KeeNsoliton, Wanga}. In order the check the convergence of the SRM routine we develop for the KEE, we used these analytically derived results as benchmark problems. We now turn our attention to the case of nonzero optical potential.

\subsection{Self-Localized Soliton Solutions of the KEE for Nonzero Optical Potential}

\begin{figure}[htb!]
\begin{center}
   \includegraphics[width=3.4in]{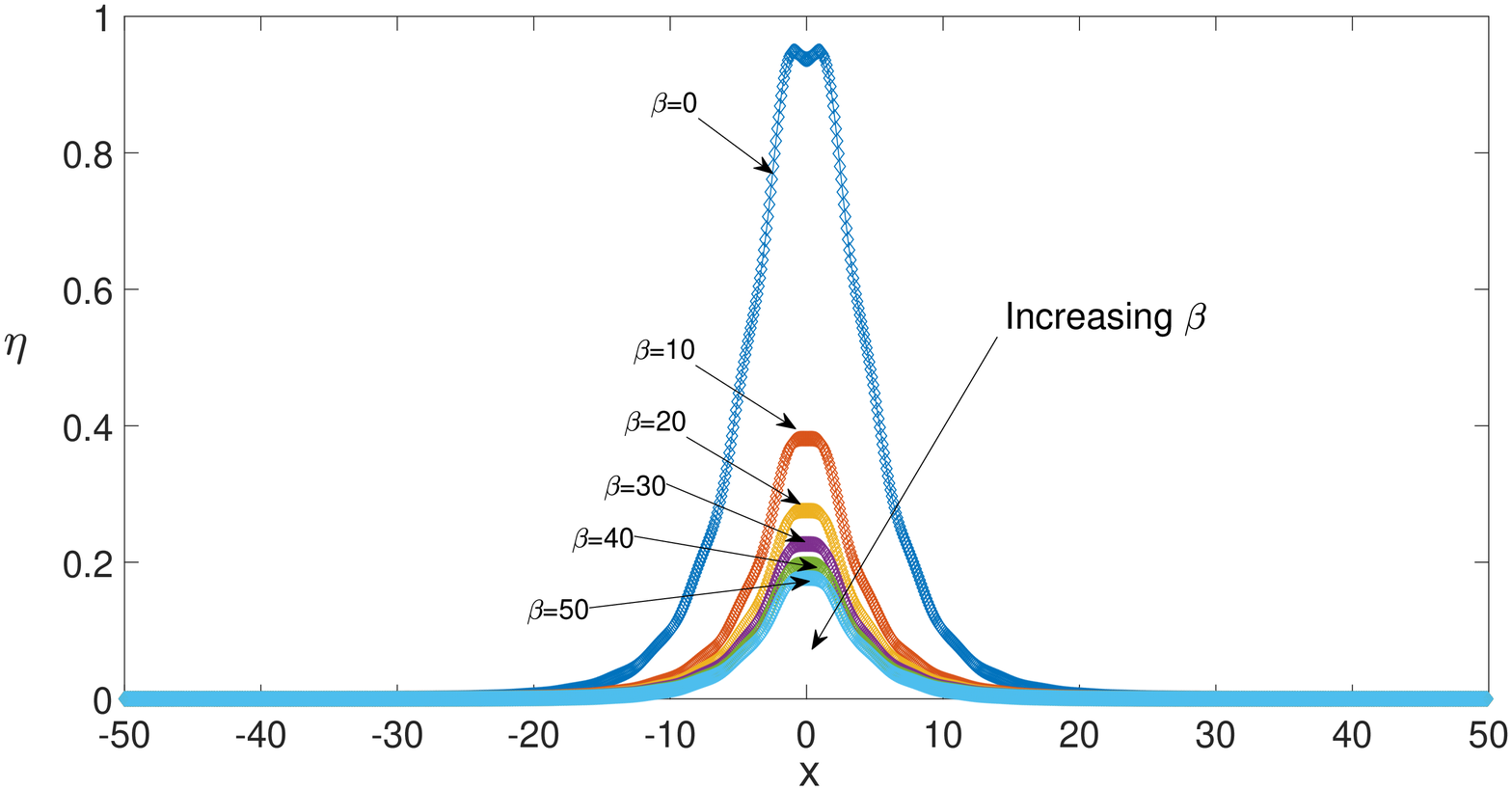}
  \end{center}
\caption{\small Self-localized soliton solution of the KEE with cubic-quintic-Raman nonlinearity for $I_o=4$ as a function of $\beta$.}
  \label{fig4}
\end{figure}

Secondly, we concentrate on the KEE with an optical potential in the form of $V=I_o \cos^2(x)$. The nonlinear term for this case is again cubic-quintic-Raman nonlinearity, $N(\left| \zeta \right|^2)=2 \left| \zeta \right|^2+\beta^2 \left| \zeta \right|^4-2 \beta i  \left( \left| \zeta \right|^2 \right)_x $ which is used in Eq.~(\ref{eq01}). The parameters of the computations presented in Figure~\ref{fig4} are selected as $p=10, \mu=1, I_0=4$. Convergence criteria is selected as $\alpha$ to be less than $10^{-15}$ as before. Checking Figure~\ref{fig4}, one can observe that SRM converges and self-localized solutions of the KEE under the effect of periodic potential exists. Additionally, as the $\beta$ term grows, the power of the solutions decreases.
 
\begin{figure}[htb!]
\begin{center}
   \includegraphics[width=3.4in]{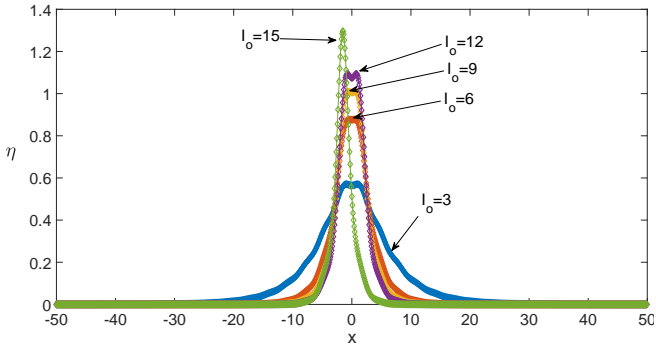}
  \end{center}
\caption{\small Self-localized soliton solution of the KEE with cubic-quintic-Raman nonlinearity for $\beta=2$ as a function of $I_0$.}
  \label{fig5}
\end{figure}

The effect of the potential amplitude,  $I_0$, on existence of self-localized solutions is investigated in Figure~\ref{fig5}. The parameters for this case are selected $p=10, \mu=1, \beta=2$. Scanning a range of $I_0=0-100$, only self-localized solutions are observed within the interval of $I_0=2.2-16.7$ and some of them are depicted in Figure~\ref{fig5}. As $I_0$ increases, the soliton amplitude and its power increase.

\begin{figure}[htb!]
\begin{center}
   \includegraphics[width=3.4in]{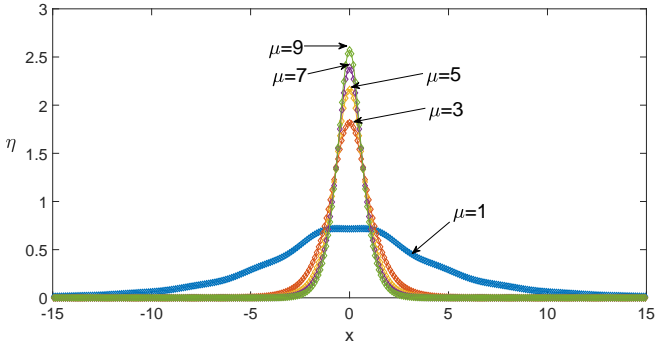}
  \end{center}
\caption{\small Self-localized soliton solution of the KEE with cubic-quintic-Raman nonlinearity for $I_0=4, \beta=2$ as a function of $\mu$.}
  \label{fig6}
\end{figure}

In our simulations with  KEE having cubic-quintic-Raman nonlinearity and periodic potential, we have observed no self-localized solitons for $\mu=[0-1[$, however for a wide range of $\mu=[1-100]$ we have observed self-localized solitons. In order to illustrate the effect of the soliton eigenvalue on the existence and shapes of self-localized solutions, we depict Figure~\ref{fig6}. Clearly, as the soliton eigenvalue grows, so does the amplitude of the self-localized solitons.

It is practically important to discuss the stability of the self-localized solutions of the KEE with cubic-quintic-Raman nonlinear term presented in the figures above. For a soliton to be stable two conditions must hold true. The first one is a necessary condition known as the slope condition, $dP(\mu)/{d\mu}<0$, as first derived by Vakhitov and Kolokolov \cite{VakhitovStability, SivanStability}. In here $P=\int \left| \zeta \right|^2 dx$ is the soliton power. The second condition is the spectral condition. To formulate the spectral condition, we can use the operator as $L_+= -\Delta + V -1[N(\eta)-2\eta^2 N'(\eta)]-\mu$ for the KEE problem that we analyze \cite{WeinsteinStability, SivanStability}. In here $\Delta$ is the diffraction term also known as the Laplacian, $V$ is the periodic potential and 
$N(\eta)$ can be two different nonlinear terms we discussed above, cubic-quintic-Raman term or its saturable version. For the stability, in addition to the slope condition, $L_+$ should have at most one eigenvalue which should be nonzero \cite{SivanStability}. A more comprehensive discussion of soliton instability can be seen in \cite{SivanStability}.

\begin{figure}[htb!]
\begin{center}
   \includegraphics[width=3.4in]{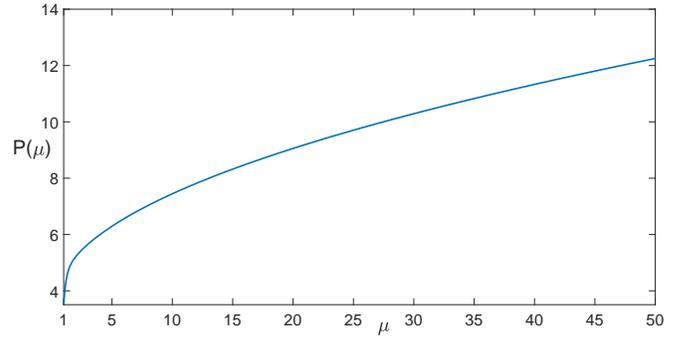}
  \end{center}
\caption{\small Power of the self-localized solutions of KEE with cubic-quintic-Raman nonlinearity for $I_0=4$ as a function of soliton eigenvalue $\mu$.}
  \label{fig7}
\end{figure}

The soliton eigenvalue vs soliton power is depicted in Figure~\ref{fig7} for $I_0=4, \beta=2$. Within the range of $\mu=[1-100]$ where the self-localized solitons of the KEE with cubic-quintic-Raman nonlinearity exist, the soliton power is increasing as soliton eigenvalue increases. Therefore, since the necessary slope condition is not satisfied, we can conclude that self-localized soliton solutions of the KEE with cubic-quintic-Raman nonlinearity are not stable, at least for the ranges of parameters considered above.

Additionally, in our simulations with KEE having cubic-quintic-Raman nonlinearity under the effect of periodic potential, we observe that initial conditions in the form of multi-Gaussians evolve into single hump solitons for the range $I_0=2.2-16.7$ and they are also unstable.

\begin{figure}[htb!]
\begin{center}
   \includegraphics[width=3.4in]{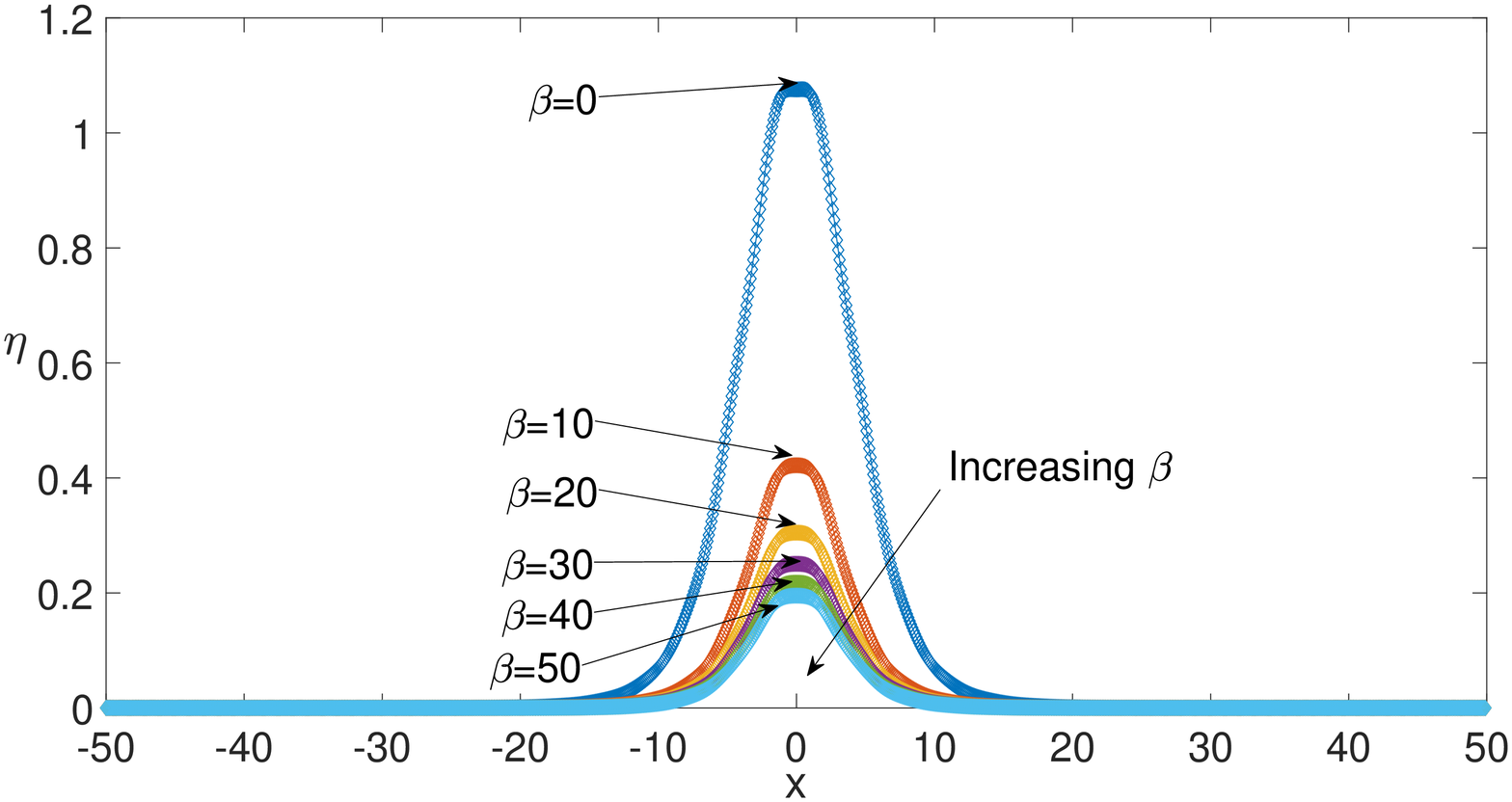}
  \end{center}
\caption{\small Self-localized soliton solution of the KEE with saturable cubic-quintic-Raman nonlinearity for $I_o=1$ as a function of $\beta$.}
  \label{fig8}
\end{figure}

\begin{figure}[htb!]
\begin{center}
   \includegraphics[width=3.4in]{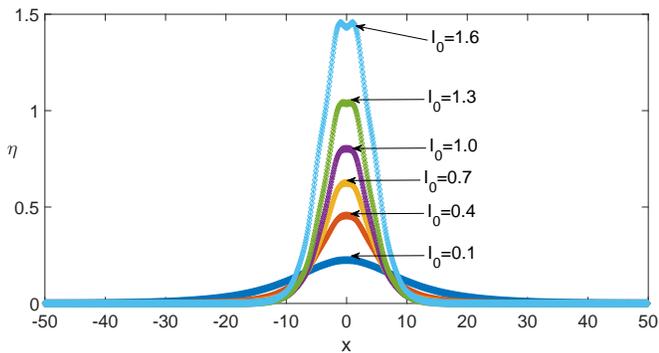}
  \end{center}
\caption{\small Self-localized soliton solution of the KEE with saturable cubic-quintic-Raman nonlinearity for $\beta=2$ as a function of $I_0$}
  \label{fig9}
\end{figure}

Next, we turn our attention to the KEE-like equation given in Eq.~(\ref{eq01}) for a nonzero optical potential of $V=I_o \cos^2(x)$ and the saturable cubic-quintic-Raman nonlinear term given as $N(\left| \zeta \right|^2)=-1/(1+2 \left| \zeta \right|^2+\beta^2 \left| \zeta \right|^4-2 \beta i  \left( \left| \zeta \right|^2 \right)_x)$ for which the iteration formula becomes Eq.~(\ref{eq09}). Calculations for this case are performed for $p=2$.  Within the range of $I_0=0-1.7$, self-localized solitons of the KEE like equation are observed as depicted in Figure~\ref{fig8} and Figure~\ref{fig9}. These solitons are similar to the solitons of the NLSE with saturable nonlinearity observed in practice and given in \cite{Segev}.

Checking Figure~\ref{fig8} and Figure~\ref{fig9}, one can conclude that soliton amplitude and power decrease as $\beta$ increases; and they increase as $I_0$ increases.

\begin{figure}[htb!]
\begin{center}
   \includegraphics[width=3.4in]{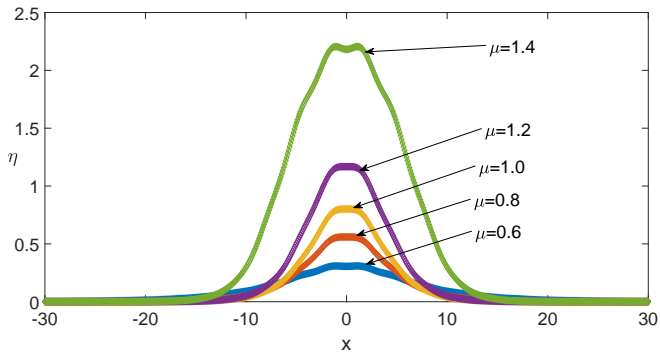}
  \end{center}
\caption{\small Self-localized soliton solution of the KEE with saturable cubic-quintic-Raman nonlinearity for $I_0=1, \beta=2$ as a function of $\mu$.}
  \label{fig10}
\end{figure}

\begin{figure}[htb!]
\begin{center}
   \includegraphics[width=3.4in]{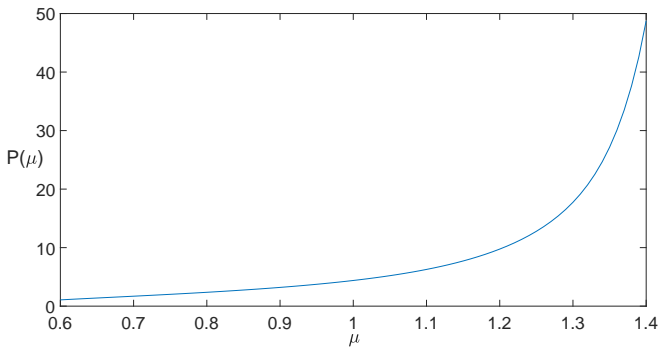}
  \end{center}
\caption{\small Power of the self-localized solutions of KEE with saturable cubic-quintic-Raman nonlinearity for $I_0=1$ as a function of soliton eigenvalue $\mu$.}
  \label{fig11}
\end{figure}

In order to check the stability of the solitons, we parametrize the soliton eigenvalue and present the corresponding results in Figure~\ref{fig10} and Figure~\ref{fig11}. As before, as these figures clearly indicate, the slope condition of Vakhitov and Kolokolov necessary for the soliton stability is not satisfied, thus it is possible to conclude that self-localized solutions of the KEE like equation are unstable, at least for the parameter ranges considered. Since the necessary slope-condition is not satisfied for all of the cases we considered, the test of the spectral condition was not needed to judge the stability of the self-localized solitons. However, under different photo-refractive media having different potential types, the results may change and the spectral condition may be needed to tested analytically and/or numerically.

\section{\label{sec:level1}Conclusion and Future Work}

In this paper we have numerically analyzed the 1D self-localized solutions of the Kundu-Eckhaus equation in nonlinear waveguides using the spectral renormalization method and have compared our findings with those solutions of the nonlinear Schr\"{o}dinger equation. We have used the analytically derived single, dual and $N$-soliton solutions of the Kundu-Eckhaus equation for the case with zero optical potential, i.e. $V=0$, as benchmark problems to test the accuracy SRM routine we developed for it.  We have showed that self-localized solutions of the Kundu-Eckhaus equation with cubic-quintic-Raman nonlinearity do exist, at least for some range of parameters, for $V=I_o \cos^2(x)$ type potentials. Additionally we have also showed that self-localized solutions of the KEE with saturable cubic-quintic-Raman nonlinearity under the effect of periodic potentials do exist for some range of parameters as dicussed in the text. However for both of these two cases, the necessary slope condition for the soliton stability is not satisfied, thus self-localized solutions obtained by the spectral renormalization method were found to be unstable. Our results can be used for studying the propagation and stability characteristics of the femto-pulses in nonlinear optics and physics and in nonlinear wave blocking problems in water wave mechanics.

\end{document}